\documentclass[10pt,conference]{IEEEtran}
\IEEEoverridecommandlockouts
\usepackage[utf8]{inputenc}
\usepackage{graphicx} % Allows including images
\usepackage{booktabs} % Allows the use of \toprule, \midrule and \bottomrule in tables
\usepackage{amsmath}
\usepackage{bm}
\usepackage{gensymb}
\usepackage{physics}
\usepackage{mathtools, nccmath}
\usepackage{algorithm}
\usepackage{algpseudocode}
\usepackage{setspace}
\usepackage{amsfonts}
\usepackage{amssymb}
\usepackage{bbm}
\usepackage{cite}
\usepackage{lipsum}
\usepackage{authblk}
\usepackage{adjustbox}

\usepackage[left=0.625in, right=0.625in, top=0.75in, bottom=1in]{geometry}

\title{An ODMA-Based Unsourced Random Access Scheme with a Multiple Antenna Receiver}
\author[*]{Mert Ozates \thanks{Mert Ozates' work is supported by the Federal Ministry of Education and Research of Germany in the programme of “Souverän. Digital. Vernetzt” joint Project 6G-RIC, project identification number: 16KISK026. }}
\author[**]{Mohammad Kazemi}
\author[***]{Tolga M. Duman}
\affil[*]{IHP - Leibniz Institute for High Performance Microelectronics, 15236 Frankfurt (Oder), Germany \protect\\ Email: oezates@ihp-microelectronics.com}
\affil[**]{Department of Electrical and Electronic Engineering, Imperial College London  \protect\\ Email: mohammad.kazemi@imperial.ac.uk}
\affil[***]{Dept. of Electrical and Electronics Engineering, Bilkent University, Bilkent 06800, Ankara, Turkey \protect\\Email:  duman@ee.bilkent.edu.tr}

\date{}

\begin{document}

\maketitle
\thispagestyle{empty}
\pagestyle{empty} 

\begin{abstract}
    We investigate the unsourced random access scheme assuming that the base station is equipped with multiple antennas, and propose a high-performing solution utilizing on-off-division multiple access. We assume that each user spreads its pilot sequence and polar codeword to the pilot and data parts of the transmission frame, respectively, based on a transmission pattern. The iterative receiver operation consists of pilot and pattern detection followed by channel vector and symbol estimation, polar decoding, and successive interference cancellation. Numerical findings demonstrate that the proposed scheme has superior performance compared to the state-of-the-art in various antenna settings.
\end{abstract}

%We assume that the users send a pilot sequence followed by their polar codeword that is spread to the rest of the transmission frame based on a transmission pattern.

%by dividing the transmission frame into pilot and data parts and 
%solution in the data part.
%performs superior
%random pattern

\begin{IEEEkeywords}
Unsourced random access, on-off division multiple access, massive MIMO.
\end{IEEEkeywords}

\section{Introduction}

Next-generation wireless communication systems require high reliability, high connection density, and low latency. This makes massive machine-type communications (mMTC) \cite{mMTC} a key feature of 5G and beyond 5G (B5G) systems. In mMTC, a massive number of devices (e.g., millions of devices per $\text{km}^2$) with low transmit power and short payloads sporadically communicate with a base station (BS) without any coordination, i.e., only a small fraction of them are active at any given time.

%Next generation systems requires .. makes mmTC.. refers massive number of machine type devices low energy comput. power sporadic strict delay req. Conventional grant-based approaches -- delay signalling overhead, sourced grant-free -- insufficient resources , Polyanskiy introduces , information theoretic formulation advanced form of grant-free random access?

%To address the communication of these devices
%these systems
%Due to the stringent delay requirements, the payloads are short, for instance, they are in the order of a few hundred bits.

Conventional grant-based multiple access approaches where the BS assigns fixed resources (time, frequency, code, etc.) to each user become infeasible in mMTC applications such as the Internet-of-things (IoT) due to the excessive delay and signaling overhead of scheduling a massive number of users. To address this problem, a new grant-free random access paradigm called unsourced random access (URA) is introduced in \cite{polyanskiy}. In URA, the devices share the same codebook; hence, the user identities are removed, which allows for an arbitrarily large number of total users. The receiver aims to recover a list of the transmitted messages regardless of the user identities, and the per-user probability of error (PUPE) is adopted as the main error metric.

% The system operation only depends on the number of active users,
% and the system operation becomes independent of total number of them
% are effective for a moderate number of users, however, they

In \cite{polyanskiy}, URA over a real-valued Gaussian multiple access channel (MAC) is studied, and a random coding achievability bound quantifying the required energy-per-bit for a given active user load is derived. Several practical transmission schemes are proposed to approach this bound based on dividing the transmission frame into slots \cite{vem,duman}, coded compressed sensing \cite{comp2}, or spreading the channel-coded bits by random signatures \cite{pradhan,javad}. Furthermore, the on-off division multiple access (ODMA) \cite{odma} where a small fraction of the frame is utilized by each user is applied to the URA set-up in \cite{odma1,odma2}, offering a promising performance with low complexity.

The fading MAC, which is a more practical model compared to the Gaussian MAC, is widely studied in URA, where a single-antenna receiver is considered either over flat \cite{fading1,fading4} or frequency-selective fading channels \cite{ozates1}. Besides, the spatial multiplexing gains of massive multiple-input multiple-output (MIMO) technology are utilized to cope with the high multiuser interference in URA. As a result, many recent works on URA examine the scenario with a massive MIMO receiver. Some typical methods in this framework are coded compressed sensing combined with non-Bayesian activity detection \cite{mimo1}, tensor-based modulation \cite{mimo3}, and the employment of pilots in a part of slot or frame for user separation \cite{twc,orth,fasura2}.

%where the BS is equipped with a massive number of antennas.
%Besides, the potential of massive multiple-input-multiple-output (MIMO) technology is utilized to cope with the high multiuser interference in URA thanks to the spatial multiplexing gains.
%where the transmitted signals of the users are attenuated

In this paper, we tackle the problem of URA with multiple antennas at the BS and develop an energy-efficient solution based on ODMA. In our proposed scheme, we assume that the channel frame is divided into the pilot and data parts, and the transmission in both parts is performed in an ODMA manner. Namely, each user randomly and independently selects a pilot sequence from a common codebook to transmit in the active transmission indices of the pilot part determined by a pilot transmission pattern. In the data part, a polar encoded and quadrature phase shift keying (QPSK) modulated sequence is sent based on a data transmission pattern determining the locations of the codeword elements in the data part of the frame. The transmitted messages are recovered by first detecting the active pilot sequences and transmission patterns using the generalized orthogonal matching (gOMP) algorithm and then employing linear minimum mean square (LMMSE) channel estimation, maximal ratio combining (MRC), single-user decoding, and successive interference cancellation (SIC) iteratively. We demonstrate the effectiveness of the proposed solution via numerical examples, showing that it outperforms the state-of-the-art for multiple antenna configurations, including massive MIMO. For instance, it can decrease the required energy-per-bit up to 2.5 dB for massive MIMO and 1.5 dB for the multiple antenna case compared to the state-of-the-art, and it can support up to 1400 active users in the massive MIMO scenario. Note that our proposed scheme is similar to that of \cite{odma2}. However, we generalize the scheme in \cite{odma2} by equipping the receiver side with multiple antennas (while \cite{odma2} assumes a single antenna receiver) and employing ODMA in the pilot part of the frame as well, and modify the related algorithmic steps accordingly.

%However, we generalize it by employing ODMA in the pilot part of the frame as well and modify the related algorithmic steps accordingly, and we consider the multiple antenna configurations at the receiver side while \cite{odma2} only examines the single antenna scenario.

% For instance, it can support up to 1400 active users in the massive MIMO scenario, and decreases the required energy-per-bit for a given PUPE up to 2.5 dB for massive MIMO, and 1.5 dB for the multiple antenna case compared to the state-of-the-art.

%and multiple antenna regimes, and that it
%including in both 
%hence, it is quite scalable as well.

The rest of the paper is organized as follows. We introduce the system model in Section \ref{system}, and give the details of the proposed solution in Section \ref{proposed}. We present a numerical evaluation of our scheme in Section \ref{numerical}, and give some conclusions in Section \ref{conclusion}.

%We present a set of numerical results in Section \ref{numerical}, and conclude the paper in Section \ref{conclusion}.

%At the receiver side,
%In multiple access scenarios, the devices simultaneously share a communication medium to transmit their data. In the case of small number of users, conventional multiple access techniques where fixed time, frequence or code resources are assigned to users are 
%are detected
% the message bits are recovered by 

\section{System Model} \label{system}

We consider a massive random access scenario where $K_T$ users sporadically communicate with a common BS, i.e., only $K_a$ of them are active at a given time where $K_a \ll K_T$. The active users transmit $B$ bits of information over a quasi-static fading channel to a  BS with $M$ antennas through a channel frame of length $n$. The received signal at the BS is

%can be written as

%multiple-antenna
%to obtain their transmitted signal, and transmit it 
%where they encode their message bits using the same codebook.
%We consider a massive random access scenario where $K_a$ active users out of $K_T$ users ($K_a \ll K_T$) transmit $B$ bits of information over a quasi-static fading channel to a multiple-antenna BS.
%Assuming the absence of synchronization errors, 

\begin{equation}
    \mathbf{Y} =  \sum\limits_{i =1 }^{K_a} \mathbf{x}_i \mathbf{h}_i + \mathbf{Z},
\end{equation}

\noindent where $\mathbf{x}_i \in\mathbb{C}^{n \times 1}$ is the transmitted signal of the $i$-th user, $\mathbf{h}_i \in\mathbb{C}^{1 \times M}$ is the channel vector of the $i$-th user consisting of independent and identically distributed (i.i.d.) elements that are constant throughout the transmission frame, $\mathbf{Y} \in\mathbb{C}^{n \times M}$ is the received signal, and $\mathbf{Z} \in\mathbb{C}^{n \times M} $ is the circularly symmetric complex additive white Gaussian noise (AWGN) with i.i.d. zero mean elements. 

%and $M$ is the number of the BS antennas. 

\begin{figure}
    \centering
    \hspace*{-5mm}
     \includegraphics[scale = 0.38]{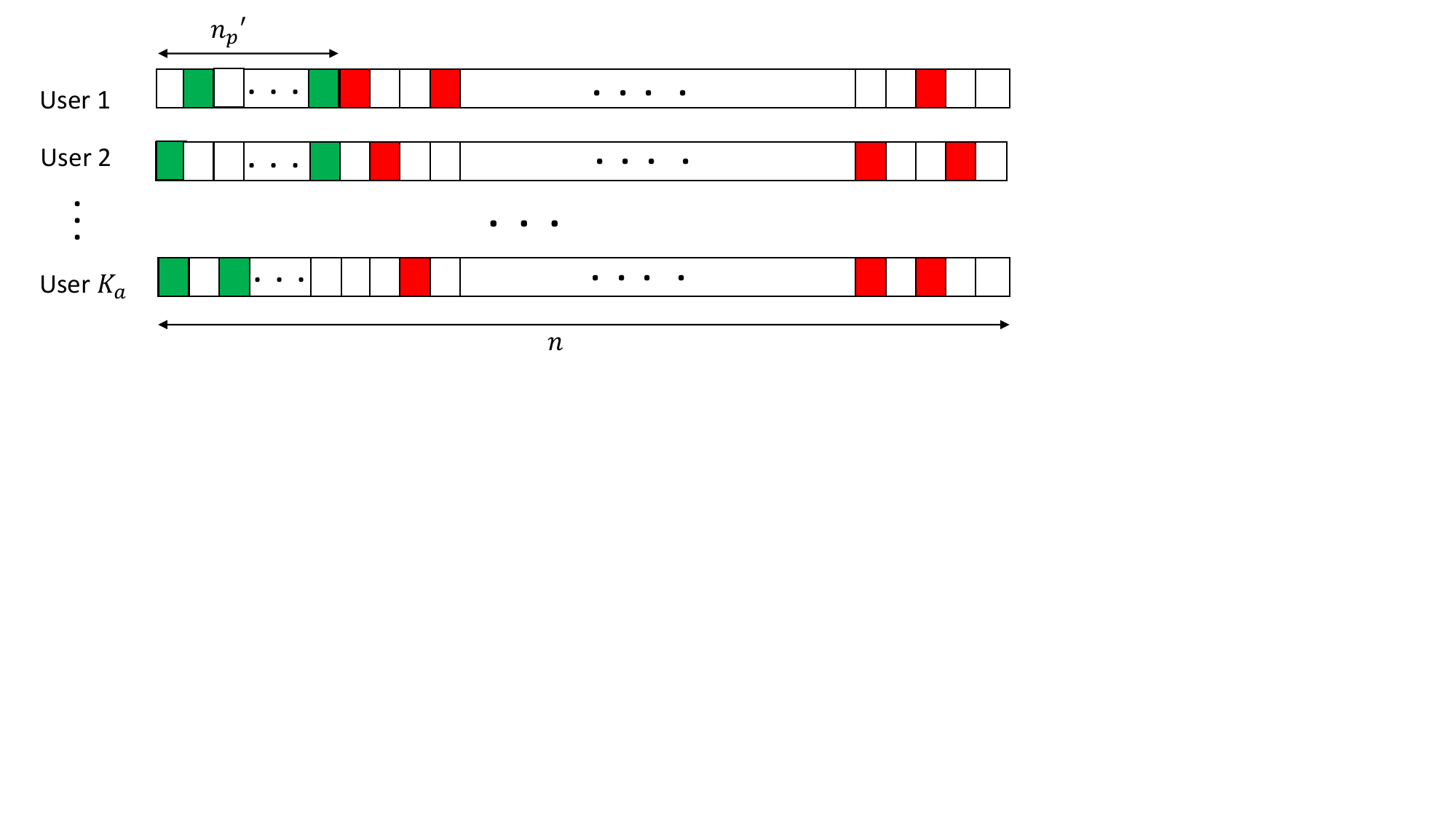}
     \vspace*{-44mm}
    \caption{The proposed transmit signal structure of multiple users. The utilized time instances in the pilot and data parts are shown by green and red boxes, respectively.}
    \label{figsignal}
\end{figure}

In our proposed model, each user transmits a pilot signal of length $n_p$ in the first $n_p'$ time instances, and utilizes $n_d$ time instances out of the remaining $n - n_p'$ ones for the data transmission as depicted in Fig. \ref{figsignal}, which illustrates the ODMA idea. The pilot and data signals have an average power constraint of $P_p$ and $P_d$, respectively, hence, the required energy-per-bit is

%of the system becomes

\begin{equation}
    \frac{E_b}{N_0} = \frac{n_pP_p + n_dP_d}{BN_0},
\end{equation}

\noindent where $N_0$ is the noise variance, i.e., the variance of each element of $\mathbf{Z}$. The purpose of the receiver is to generate a list of the decoded messages $\mathcal{L} = \{ \mathbf{\hat{m}}_1, \dots, \mathbf{\hat{m}}_{K_a} \}$. The system PUPE $P_e$ is evaluated in terms of the misdetection and false alarm probabilities $P_{\text{md}}$ and $P_{\text{fa}}$, respectively, which can be calculated as

%of the

\begin{equation}
  P_{\text{md}} = \frac{1}{K_a} {\mathbb{E}\bigg[\sum\limits_{i \in \mathcal{K}_a} \mathbbm{1}_{\{\mathbf{m}_i \notin \mathcal{L}\}}\bigg]},
 % P_{\text{md}} =  \frac{\mathbb{E}\bigg[\sum\limits_{k \in \mathcal{K}_a} \mathbbm{1}_{\{\mathbf{m}_k \notin \mathcal{L}\}}\bigg]}{K_a},   
\end{equation}

\begin{equation}
   P_{\text{fa}} = \mathbb{E} \left[\frac{\abs{\mathcal{L} \setminus  \{\mathbf{m}_i: i \in \mathcal{K}_a \}}}{\abs{\mathcal{L}}}\right],
\end{equation}

\noindent where $\mathbf{m}_i$ is the message of the $i$-th user, $\mathcal{K}_a$ is the set of active users, $\mathbbm{1}_{\{\ \hspace{-1mm} \cdot \}}$ is the indicator function, $\abs{\cdot}$ denotes the cardinality of a set, and the expectation is taken over the randomness of the fading and noise processes. The PUPE can be obtained by summation of these two probabilities, i.e., $P_e = P_{\text{md}} + P_{\text{fa}}$. 

%with the pilot and data parts

\section{Proposed Scheme} \label{proposed}

\subsection{Encoding}

We assume that each user divides its message bits into two parts, $\mathbf{m}_p$ and $\mathbf{m}_d$, respectively. The first part $\mathbf{m}_p$ determines the pilot and pattern indices of the users. For the transmission in the pilot part, a user picks the $ind(\mathbf{m}_p) = \text{dec}(\mathbf{m}_p) + 1$-th column of the common non-orthogonal pilot matrix $\mathbf{A} \in\mathbb{C}^{n_p \times N}$ as its pilot sequence, where $N = 2^{B_p}$, $B_p$ is the length of $\mathbf{m}_p$, and $\text{dec}(.)$ denotes binary-to-decimal conversion. The columns of $\mathbf{A}$ are normalized to have a Euclidean norm of $\sqrt{n_pP_p}$ to satisfy the power constraint. The selected pilot sequence is then transmitted in the first $n_p'$ time instances of the transmission frame based on a pilot transmission pattern, that is the $ind(\mathbf{m}_p)$-th column of the pilot transmission pattern matrix $\mathbf{P}_{\text{pilot}} \in \mathbb{R}^{n_p' \times N}$ with $n_p$ non-zero elements (activity indices) in each column. The indices of non-zero elements show the time instances where the pilot symbols will be located, while the rest are the idle indices. Note that we have dropped the user index for ease of presentation. The pilot part of the received signal can be written as

%Note that also determined by the first $B_p$ bits.

%that is also determined by the first $B_p$ bits.

\begin{equation}
    \mathbf{Y}_p =  \sum\limits_{i =1 }^{K_a} {s}_p (\mathbf{a}_i) \mathbf{h}_i + \mathbf{Z}_p,
\end{equation}

\noindent where $\mathbf{Y}_p \in\mathbb{C}^{n_p' \times M}$, $\mathbf{a}_i \in\mathbb{C}^{n_p \times 1}$ is the pilot signal of the $i$-th user, ${s}_p (\mathbf{a}_i) \in \mathbb{C}^{n_p' \times 1}$ is the pilot part of the transmitted signal of the $i$-th user, $s_p(.)$ is a mapper distributing the pilot signal to the pilot part of the frame, and $\mathbf{Z}_p \in\mathbb{C}^{n_p' \times M}$ is the sub-matrix consisting of the first $n_p'$ rows of $\mathbf{Z}$. 

%that are called

\begin{figure}
    \centering
    \hspace*{7mm}
     \includegraphics[scale = 0.63]{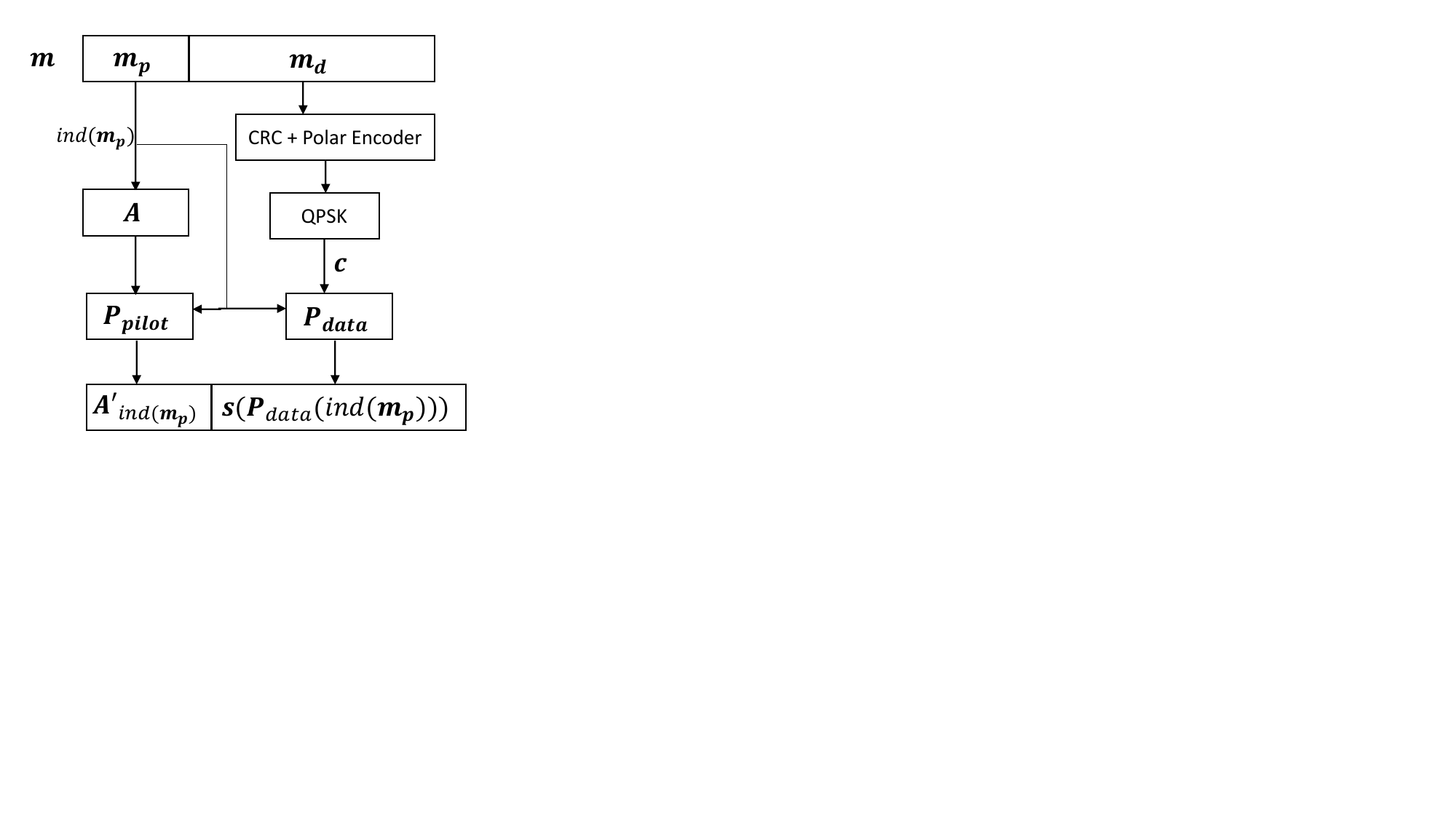}
     \vspace*{-56mm}
    \caption{Encoding process for a user in the proposed scheme.}
    \label{figencoder}
\end{figure}

The second part $\mathbf{m}_d$ is encoded by a $(n_c, B-B_p + r)$ polar code and modulated by QPSK to be transmitted in the last $n - n_p$ instances, where $n_c$ is the code length and $r$ is the number of cyclic redundancy check (CRC) bits. CRC is used to ensure that the decoded sequence is a valid codeword to avoid error propagation in an iterative decoder. The encoded and modulated sequence is then distributed to the data part of the frame based on a data transmission pattern, that is the $ind(\mathbf{m}_p)$-th column of the data pattern matrix $\mathbf{P}_{\text{data}} \in \mathbb{R}^{(n-n_p') \times N}$ with $n_d$ non-zero elements in each column, denoting the locations of the codeword elements. The received signal in the data part can be written as

%The indices of non-zero elements show the indices where the data symbols will be located, while the rest are the idle indices. 

\begin{equation}
  %  \mathbf{Y}_d = \sum\limits_{i =1 }^{K_a} \mathbf{s}_i(\mathbf{p}_i) \mathbf{h}_i + \mathbf{Z}_d,
     \mathbf{Y}_d = \sum\limits_{i =1 }^{K_a} {s}_d(\mathbf{c}_i) \mathbf{h}_i + \mathbf{Z}_d,
\end{equation}

\noindent where $\mathbf{Y}_d \in\mathbb{C}^{(n - n_p') \times M}$, ${s}_d(\mathbf{c}_i) \in \mathbb{C}^{(n - n_p') \times 1}$ is the transmitted data signal of the $i$-th user, $s_d(.)$ is a mapper that distributes the modulated polar codeword of the $i$-th user $\mathbf{c}_i$ with $n_d$ non-zero elements in $\{ \sqrt{P_d/2} (\pm 1 \pm j)\}$ to the data part, and $\mathbf{Z}_d$ is the matrix of the last $n - n_p'$ rows of $\mathbf{Z}$. The encoding procedure is illustrated in Fig. \ref{figencoder}. 

%based on its data transmission pattern

%data part of the transmitted
%of the frame
%Note that if all of the indices of the pilot part are active, the model simplifies to the one in \cite{odma2}. 
%where $ind(\mathbf{m}_p) = \text{dec}(\mathbf{m}_p) + 1$ is the pilot/patten index corresponding to $\mathbf{m}_p$. 

%$\mathbf{p}_i$,

%to a s constructed according to its transmission pattern $\mathbf{p}_i$ by the mapper ${s}(.)$
%Also, the transmit signal structure of the active users is depicted in Fig. \ref{figsignal}, which illustrates the ODMA idea.

\subsection{Decoding Procedure} \label{decode}

\subsubsection{Joint Active Pilot and Pattern Detection} 

In order to decode the transmitted message sequences, the active pilot sequences should be detected first. Since $K_a \ll N$, this is essentially a compressed sensing problem, and we employ the gOMP algorithm \cite{gomp} to detect the active pilots. 

Since the pilot sequences are sent in the active indices of the pilot part, the pilot signal of any specific user occupies a fraction of the pilot part. Hence, the pilot codebook $\mathbf{A}$ cannot be directly used for activity detection. To solve this issue, we extend the pilot codebook $\mathbf{A}$ using the pilot transmission patterns. Namely, we define an extended pilot codebook $\mathbf{A'} \in \mathbb{C}^{n_p' \times N}$ and construct the $i$-th column of it by putting the elements of the $i$-th column of $\mathbf{A}$ to the activity indices dictated by the $i$-th column of $\mathbf{P}_{\text{pilot}}$, while the rest of the column elements are zero. We then utilize $\mathbf{A'}$ as an effective pilot codebook to detect the active pilots and pilot patterns jointly, as $\mathbf{A}'$ has the information of pilot transmission patterns as well.

%Due to its construction, the extended pilot codebook has the information of both the pilot sequences and pilot transmission patterns, hence, by detecting the active columns of it both the active pilot and transmission pattern indices will be identified.

%we put the elements of each pilot sequence to the active indices dictated by the pilot transmission pattern of the extended pilot codebook $\mathbf{A}' \in \mathbb{C}^{n_p' \times N}$ and use it as the pilot codebook for activity detection. 

In the first step of each gOMP iteration, the correlation of the effective pilots and the received signal is obtained as 

%generalized orthogonal matching pursuit 

\begin{equation}
    \mathbf{C} = \mathbf{A'}^H \mathbf{Y}_p^{(k)},
    \label{eqomp}
\end{equation}

\noindent where $\mathbf{C} \in\mathbb{C}^{N \times M}$ and $\mathbf{Y}_p^{(k)}$ is the residual pilot signal at the $k$-th gOMP iteration initialized as $\mathbf{Y}_p^{(1)} = \mathbf{Y}_p$. The Euclidean norm of each row of $\mathbf{C} $ gives the decision metric for the corresponding pilot. We then add the indices of the $ \lceil \frac{K_a + \Delta}{n_\text{OMP}} \rceil$ elements having the maximum decision metric to the output list of gOMP, where $n_\text{OMP}$ is the number of gOMP iterations and $\Delta$ is a small positive integer. In the second step, the effect of the detected pilots is subtracted as

%using the projection of them on the received pilot as

%by taking the projection of them  using a linear minimum mean square error (LMMSE) solution and subtract their effect from the residual signal as follows 

%In the second step, we calculate the temporary channel vector estimates corresponding to the detected pilots using a linear minimum mean square error (LMMSE) solution and subtract their effect from the residual signal as follows 

%\begin{equation}
  %  \hat{\mathbf{H}}^{(k)} = (\mathbf{A}_{\hat{{I}}^{(k)}}^H\mathbf{A}_{\hat{{I}}^{(k)}} + N_0\mathbf{I}_{\abs{{\hat{I}}^{(k)}}})^{-1} \mathbf{A}_{\hat{{I}}^{(k)}}^H \mathbf{Y}_p,
%\end{equation}

\begin{equation}
   \mathbf{Y}_p^{(k + 1)} = \mathbf{Y}_p - \mathbf{A'}_{\hat{{I}}^{(k)}}(\mathbf{A'}_{\hat{{I}}^{(k)}}^H\mathbf{A'}_{\hat{{I}}^{(k)}} + N_0\mathbf{I}_{\abs{{\hat{I}}^{(k)}}})^{-1} \mathbf{A'}_{\hat{{I}}^{(k)}}^H \mathbf{Y}_p, 
    \label{eqsubt}
\end{equation}

\noindent where $\mathbf{A'}_{\hat{{I}^{(k)}}}$ is the column set of $\mathbf{A'}$ determined by $\hat{{I}^{(k)}}$, $\hat{I}^{(k)}$ is the set of detected pilot indices up to the $k$-th iteration, and $\mathbf{I}_B$ is the $B \times B$ identity matrix. The iterations continue until $n_\text{OMP}$ iterations are performed, and the output list of the detected pilots as well as pattern indices $\hat{I}$ is obtained as they are the same for a specific user since both of them are determined by the first $B_p$ bits, which are also decoded in this step.  

% with the residual signal starting (\ref{eqomp}) 
%and after the last iteration
%Furthermore, temporary channel vector estimates at the last iteration becomes estimated user channel vectors $\hat{\mathbf{H}}$.

%$\hat{\mathbf{H}}^{(k)} \in \mathbb{C}^{\abs{{\hat{I}}^{(k)}} \times M}$ is the set of temporary channel vector estimates in the $k$-th gOMP iteration

\subsubsection{Channel Vector Estimation}

Given the set of effective active pilot sequence estimates ${\mathbf{\hat{A'}}}$, we estimate the channel vectors of the active users by employing an LMMSE solution as

%linear minimum mean square error (LMMSE)
%set of
\begin{equation}
   % \mathbf{\hat{H}} = (\mathbf{A}_{\hat{{I}}}^H\mathbf{A}_{\hat{{I}}} + N_0\mathbf{I}_{\hat{K_a}})^{-1} \mathbf{A}_{\hat{{I}}}^H \mathbf{Y}_p^{(j)},
        \mathbf{\hat{H}} = (\mathbf{\hat{A'}}^H\mathbf{\hat{A'}} + N_0\mathbf{I}_{\hat{K_a}})^{-1} \mathbf{\hat{A'}}^H \mathbf{Y}_p^{(j)},
    \label{eqest}
\end{equation}

\noindent where  $\hat{K_a}$ is the cardinality of ${\hat{I}}$ and $\mathbf{Y}_p^{(j)}$ is the residual pilot signal in the $j$-th decoding iteration.

\subsubsection{Symbol Estimation and Channel Decoding}

Given the channel vector and data pattern estimates $\mathbf{\hat{H}}$ and $\mathbf{\hat{P}}_{\text{data}}$, the other data bits are decoded by first estimating user symbols by MRC and then passing them to a single-user decoder. For this purpose, using MRC, we obtain the symbol estimates of the $i$-th user by

%the symbol estimates of the $i$-th user can be obtained as 

\begin{equation}
    \hat{\mathbf{s}}_i = \mathbf{Y}_d^ {(j)} \left(\mathbf{\hat{p}}_{\text{data}} (i) \right) \mathbf{\hat{h}}_i^H,
    \label{eqmrc}
\end{equation}

\noindent where $\mathbf{Y}_d^{(j)} (\mathbf{\hat{p}}_{\text{data}} (i))$ is the set of rows of $\mathbf{Y}_d^{(j)}$ specified by the active data indices determined by $\mathbf{\hat{p}}_{\text{data}}(i)$, and $\mathbf{\hat{p}}_{\text{data}}(i)$ and $\mathbf{\hat{h}}_i$ are the data pattern and channel vector estimates of the $i$-th user, respectively. We then pass these symbol estimates to a channel decoder employing successive cancellation list decoding (SCLD), where the log-likelihood ratio (LLR) values are calculated by treating the symbol estimates as the output of a single-user channel using the fact that the users are separated by MRC. The decoded sequence is assumed to be valid and added to the output list if the CRC check is satisfied.

%if it satisfies the CRC check.

%quasi-static fading 
%the channel decoding stage
%of $\mathbf{Y}_d^{(j)}$
%that is the residual signal of $\mathbf{Y}_d$ in the $j$-th iteration

\subsubsection{Successive interference cancellation}

Due to the high multiuser interference, SIC is an essential step in URA schemes to achieve low error rates. In the case that the initially estimated channel vectors are used directly for SIC, it is performed as

\begin{equation}
      \mathbf{Y}^{(j + 1)} = \mathbf{Y}^{(j)} - \mathbf{\hat{X}}_{\hat{\mathcal{D}}}  \mathbf{\hat{H}}_{\hat{\mathcal{D}}} ,  
      \label{eqsic}
\end{equation}

\begin{figure*}
    \centering
    \hspace*{20mm}
     \includegraphics[scale = 0.65]{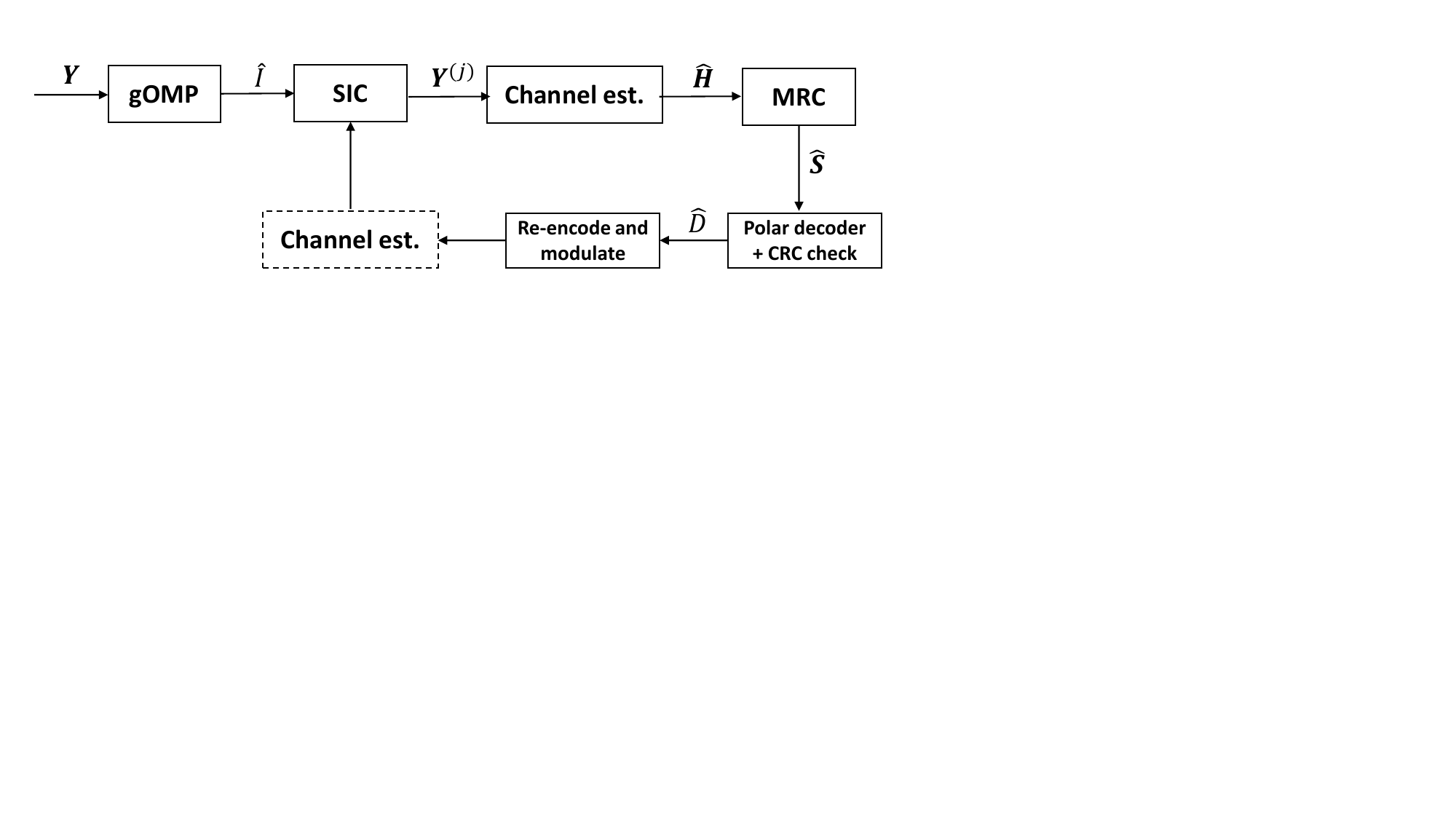}
     \vspace*{-85mm}
    \caption{Decoding process of the proposed scheme.}
    \label{figdecoder}
\end{figure*}

\noindent where $\mathbf{Y}^{(j)}$ is the residual received signal at the $j$-th iteration, $\mathcal{\hat{D}}$ is the set of the successfully decoded sequences, $\mathbf{\hat{X}}_{\hat{\mathcal{D}}}$ is the matrix including the re-constructed transmitted symbols, and $\mathbf{\hat{H}}_{\hat{\mathcal{D}}}$ is the set of channel vector estimates of the successfully decoded users. Another approach is to utilize the re-encoded and modulated decoded message sequences to re-estimate the channel vectors and perform SIC using them as

\begin{equation}
    \mathbf{\hat{H}}_{\text{re}} = (\mathbf{\hat{X}}_{\hat{\mathcal{D}}}^H\mathbf{\hat{X}}_{\hat{\mathcal{D}}} + N_0\mathbf{I}_{\abs{\mathcal{\hat{D}}}})^{-1} \mathbf{\hat{X}}_{\hat{\mathcal{D}}}^H \mathbf{Y}^{(j)},
    \label{eqreest}
\end{equation}

\begin{equation}
    \mathbf{Y}^{(j + 1)} = \mathbf{Y}^{(j)} - \mathbf{\hat{X}}_{\hat{\mathcal{D}}}  \mathbf{\hat{H}}_{\text{re}}.
    \label{eqsic3}
\end{equation}

\begin{algorithm}[t]
\caption{Decoding algorithm of the proposed scheme}\label{alg:cap}
\begin{algorithmic}[1]

\State \textbf{Input}: $\mathbf{Y}_p$, $\mathbf{Y}_d$, $\mathbf{A}$, $\mathbf{P}_{\text{pilot}}$, $\mathbf{P}_{\text{data}}$, $\Delta$, $n_{\text{max}}$, $n_{\text{OMP}}$

\State Construct the extended pilot codebook $\mathbf{A'}$ by $\mathbf{A}$ and $\mathbf{P}_{\text{pilot}}$.

\State \textbf{Activity detection}:
\For{$k=1,2, \ldots n_{\text{OMP}} $}
\State $\mathbf{C} = \mathbf{A'}^H \mathbf{Y}_p^{(k)}$.
\State Get the decision metrics by the Euclidean norm of rows of $\mathbf{C}$, take the $i_{\text{OMP}} = \left \lceil{({{K_a}} + \Delta)/n_{\text{OMP}}}\right \rceil$  indices with the maximum decision metrics and add them to ${\hat{I}}$.
%Calculate the Euclidean norm of each row of $\mathbf{C}$ to get the decision metrics,
%\State take $i_{\text{OMP}} = \left \lceil{(\frac{K_a}{V} + \Delta)/n_{\text{OMP}}}\right \rceil$ indices corresponding to the maximum elements of $\mathbf{C}$ and add them to the set $\mathcal{\hat{I}}$.
%\State If the termination condition is satisfied, go to Step 10.
%\State $ \mathbf{Y}_p^{(k + 1)} = \mathbf{Y}_p - \mathbf{A}_{\hat{{I}}^{(k)}}(\mathbf{A}_{\hat{{I}}^{(k)}}^H\mathbf{A}_{\hat{{I}}^{(k)}} + N_0\mathbf{I}_{\abs{{\hat{I}}^{(k)}}})^{-1} \mathbf{A}_{\hat{{I}}^{(k)}}^H \mathbf{Y}_p,$
\State Subtract the effect of the detected indices by (\ref{eqsubt}).
\EndFor
\State \textbf{Iterative decoding}:
\For{$j=1,2, \ldots n_{\text{max}} $}
\State Set $\mathcal{\hat{D}} = \emptyset$ , and $\hat{K}_a = \abs{\mathcal{\hat{I}}}$.
\State Channel vector estimation by (\ref{eqest}).
\For{$i=1,2,\ldots, \hat{K}_a$}
\State MRC: Estimate $\mathbf{\hat{s}}_i$ by (\ref{eqmrc}).
\State Pass $\mathbf{\hat{s}}_i$ to the polar decoder $\rightarrow$ $\mathbf{\hat{m}}_i$.
\State CRC check.
\EndFor
\If {$  \abs{\mathcal{\hat{D}}} = 0$}
\State Algorithm termination.
\EndIf
\For{i in $ \mathcal{\hat{D}}$}
\State Re-encode and modulate $\mathbf{\hat{m}}_i$ 
\EndFor
\State \textbf{SIC}:
\State Interference cancellation by (\ref{eqsic}) or re-estimation and interference cancellation by (\ref{eqreest}) and (\ref{eqsic3}).
%\If {$\mathbf{card}(\mathcal{\hat{D}}) = 0$}
\EndFor
\State $\textbf{Output}$: List of the decoded messages

%\State Estimate the channel vectors $\mathbf{\hat{H}}$ by (\ref{eqest}).
%is satisfied, $\mathcal{\hat{D}} =\mathcal{\hat{D}} \cup i$
%\While{$N \neq 0$}
%\If{$N$ is even}
   % \State $X \gets X \times X$
   % \State $N \gets \frac{N}{2}$  \Comment{This is a comment}
%\ElsIf{$N$ is odd}
   % \State $y \gets y \times X$
%    \State $N \gets N - 1$
%\EndIf
%\EndWhile

\end{algorithmic}
\end{algorithm}

Our extensive simulations show that the SIC method combined with the data-aided channel estimation is superior to one with the initially estimated channel vectors in the lower multiuser interference regimes since the non-decoded users act as interference. A comparison of these two methods is provided in Section \ref{numerical}.

%if the number of successfully decoded users is sufficient

After SIC, the pilot and pattern indices of the successfully decoded messages are removed from $\hat{I}$, which is given back to the channel estimation stage along with the residual signal. The decoding iterations continue until no valid codewords are detected in the current iteration or a predetermined number of iterations $n_{\text{max}}$ are reached. The decoding process of the proposed scheme is depicted in Fig. \ref{figdecoder}, where $\mathbf{\hat{S}}$ is the set of user symbol estimates, and a pseudo-code of the decoding process is provided in Algorithm \ref{alg:cap}.

\subsection{Complexity Analysis}

We provide a computational complexity analysis in terms of the number of multiplications per decoding iteration in this subsection. The complexity of the correlation step of gOMP in (\ref{eqomp}) is $\mathcal{O} (MNn_p)$, which is the dominant complexity of activity detection. The complexities of channel estimation and MRC are $\mathcal{O} (K_a^2 n_p)$ and $\mathcal{O} (K_aMn_d)$, respectively, and re-estimation step for SIC has a complexity  $\mathcal{O} (K_a^2 (n_p + n_d))$, dominating the complexity of our proposed scheme. Note that one can implement SCLD by only additions and subtractions \cite{polar}. The complexity of this operation is $\mathcal{O} (K_a n_L  n_c \hspace{0.5mm} \log  n_c)$, where $n_L$ is the list size.

%with a complexity of $\mathcal{O} (K_a n_L  n_c \hspace{0.5mm} \log  n_c)$, where $n_L$ is the list size.

%of the channel vectors

\subsection{Collisions}

In the URA schemes, the active users pick a pilot, preamble, or pattern sequence from a common codebook based on a part of their message bits as a typical approach. 
% This is because the total number of users is very large, making assigning an individual sequence to each user impossible. 
Therefore, when the message bits that are allocated for this purpose are the same for some different users, a collision event happens. If the channel is a Gaussian MAC, the collision of two users is recovered with a high probability \cite{pradhan}.  For the effect of collisions on the system performance in the fading MAC scenario and a detailed analysis of collision probability, one can refer to \cite{twc}.

%Note that 

%for an analysis of collision probability in URA and further details.

 %The probability of this event depends on $K_a$ and $B_p$, that is chosen such that collision of three or more users is zero.
% In the case of collision of two users, we assume that if the channel vector of one of the users has a considerably higher Euclidean norm, the message of that user will be decoded successfully \cite{twc}.

  %that is also equal to 
%and subtraction step in (\ref{eqsubt}) is and $\mathcal{O} (K_a^2 n_p)$,
%$\mathcal{O} (K_a^2 n_p)$
%he complexity of correlation step of gOMP in (\ref{eqomp}) is $\mathcal{O} (MNn_p)$ and subtraction steps in (\ref{eqsubt}) is and $\mathcal{O} (K_a^2 n_p)$, that is also equal to the complexity of LMMSE.
%Note that we also detect the transmission patterns of active users as pilot and pattern index of a user is same since it is also determined by first Bp bits.

%which allows adding multiple indices to the output list at each iteration which reduces the complexity while only one index can be added to the output list in the conventional OMP.

%column index of the common non-orthogonal pilot matrix $\mathbf{A} \in\mathbb{C}^{n_p \times N}$ picked by that user where $N = 2^{B_p}$, $B_p$ is the length of the first part, and the column index is obtained by $\text{dec}(\mathbf{m}_p) + 1$ where $\text{dec}(.)$ denotes binary-to-decimal conversion.

\section{Numerical Results} \label{numerical}

We evaluate the performance of the proposed scheme in this section. We take $n = 3200$, $B = 100$, $\epsilon = 0.05$ for $M = 50$ and $\epsilon = 0.1$ for $M = 8$, and assume Rayleigh fading channels with i.i.d. channel coefficients across the users and the antennas for a more straightforward comparison with the existing literature, where $\epsilon$ is the target PUPE. We utilize 5G polar codes and set the code length to 1024, the CRC length to 16, the list size of SCLD to 128, and $n_\text{OMP} = 4$. We employ a sub-sampled discrete Fourier transform (DFT) matrix as the pilot codebook which reduces the complexity of the correlation step of gOMP to $\mathcal{O}(M N \text{log}N)$ by converting matrix multiplications to DFT operations \cite{fengler2}. Furthermore, we assume that the active indices for each transmission pattern are uniformly and randomly generated, and choose $B_p$ such that the probability of collision of more than two users is negligible.

\begin{figure}
    \centering
    \includegraphics[width=1\linewidth]{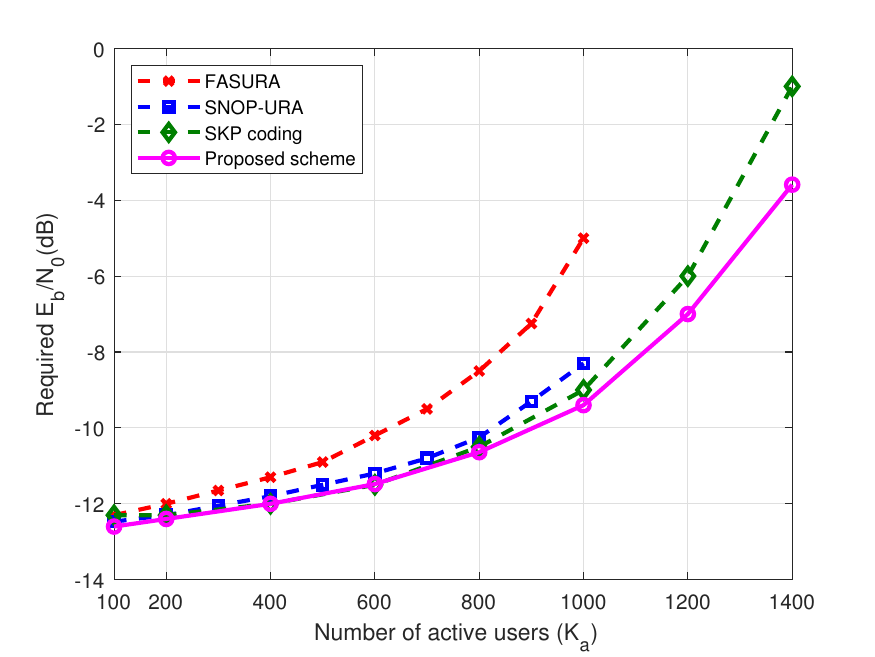}
    %\includegraphics[scale = 0.2]{fig_eff.eps}
    %\vspace*{-5mm}
    \caption{Comparison of the required $E_b/N_0$ versus the number of active users  for $M = 50$ and $P_e \leq 0.05.$}
    \label{figmassive}
   % \vspace{-3mm}
\end{figure}

We assess the energy efficiency by calculating the required $E_b / N_0$ by performing a 2D search over $P_p$ and $P_d$ and comparing it with those of some state-of-the-art solutions. In the massive MIMO regime where $M = 50$, we take $B_p =15$ for $K_a < 800$ and $B_p =16$ otherwise, and $n_p = 600$ and $n_p' = 1000$ for $ 100 \leq K_a \leq 600$, $n_p = 800$ and $n_p' = 1200$ for $K_a = 800$ and $n_p = 1000$ and $n_p' = 1300$ for $K_a > 800$ in our simulations. Also, we try both of the SIC methods discussed in Sec. \ref{decode} and get the minimum of the resulting required $E_b / N_0$ values for a given active user load. The results in Fig. \ref{figmassive} demonstrate that our proposed scheme offers the best performance compared to the state-of-the-art for $K_a \geq 800$ while it can support up to 1400 users. Namely, it outperforms FASURA \cite{fasura2} and SNOP-URA \cite{twc} by a margin of as high as 4.5 dB and 1.5 dB for $K_a \leq 1000$, respectively, and sparse Kronecker product (SKP) coding scheme in \cite{skp2} by up to 2.5 dB for $K_a \leq 1400$. Note that in the SKP coding scheme, PUPE is measured by only misdetections; hence, its performance can be slightly worsened if false alarms are also considered.

% with a maximum of 
%of the proposed scheme 
%in Fig. \ref{figmassive}
%for a target PUPE of $\epsilon = 0.05$

\begin{figure}
    \centering
    \includegraphics[width=1\linewidth]{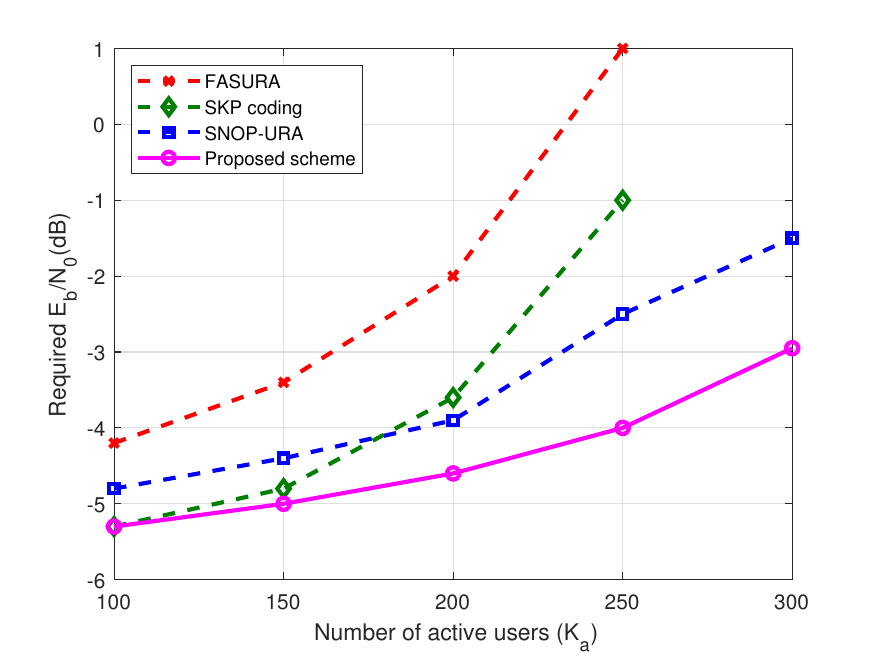}
    %\includegraphics[scale = 0.2]{fig_eff.eps}
    %\vspace*{-5mm}
    \caption{Comparison of the required $E_b/N_0$ versus the number of active users for $M = 8$ and $P_e \leq 0.1.$}
    \label{figmult}
   % \vspace{-3mm}
\end{figure}

In the multiple antenna regime where $M = 8$, the proposed scheme has excellent performance as illustrated in Fig. \ref{figmult}, where we take $n_p = 800$ and $n_p' = 1200$, and $B_p = 15$. It performs better than FASURA and SKP coding by up to 5 and 3 dB for $K_a \leq 250$, respectively, and it offers a superior performance compared to SNOP-URA by up to 1.5 dB for $K_a \leq 300$. 

%\begin{table}
%\centering
%\caption{The length of pilot sequence and pilot part for $M = 50$ for different active user loads.}
%\begin{adjustbox}{width=0.5\columnwidth,center} 
%\begin{tabular}{|c | c | c | c | c  | c|} 
% \hline
% $K_a$ & $n_p$ & $n_p'$   \\ [0.5ex] 
%\hline
% $100-600$  & 600 & 1000   \\ 
% \hline
% $800$  &  800 & 1200  \\
% \hline
% $1000-1400$ & 1000  &  1300  \\
% \hline
%\end{tabular}
%\end{adjustbox}
%\label{table:power}

%\end{table}

%of the proposed scheme and the existing schemes in the literature

\begin{figure}
    \centering
    \includegraphics[width=1\linewidth]{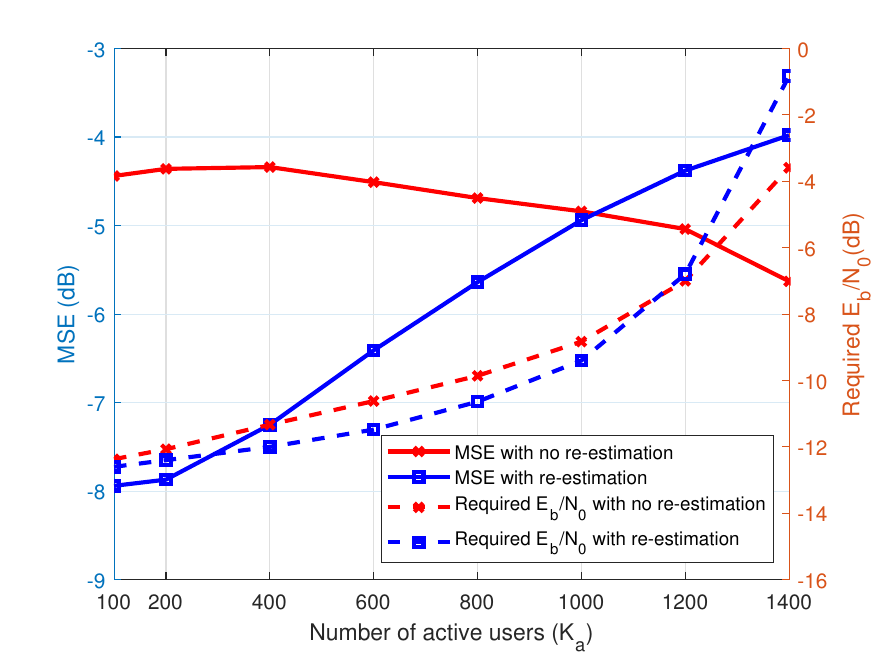}
    %\includegraphics[scale = 0.2]{fig_eff.eps}
    %\vspace*{-5mm}
    \caption{Comparison of the MSE and required $E_b/N_o$ versus number of active users for different SIC methods.}
    \label{figsic}
   % \vspace{-3mm}
\end{figure}

We also compare the mean square error (MSE) performance of candidate SIC solutions introduced in Sec. \ref{decode} normalized over the number of antennas using the same parameters in Fig. \ref{figmassive}. For simplicity, we only consider the first iteration. The MSE results in Fig. \ref{figsic} show that the data-aided channel estimation is more beneficial for most of the active user regime, however, the initial channel estimates can be used in the high multiuser interference regime instead of them. The effect of this result on system performance can also be seen in Fig. \ref{figsic}, showing that re-estimating the channel vectors using the decoded symbols provides a performance advantage for $K_a \leq 1000$, however, one should utilize the initially estimated channel vectors for $K_a > 1000$ as it provides a similar or better performance with a lower complexity.

%For $K_a = 1200$, the MSE of the re-estimation case is higher in the first iteration, however, it is possibly compensated in the subsequent iterations as the multiuser interference decreases, leading to a similar performance.
%In order to see the effect of this result on system performance, we also conduct end-to-end simulations. 
% where $K_a > 1000$
 %and calculate the MSE between the actual channel vectors and initially and re-estimated channel vectors

\section{Conclusions} \label{conclusion}

We study URA with multiple antennas at the receiver side, and propose an ODMA-based scheme where the channel frame is divided into two parts, assuming that the active users distribute their pilot and data symbols to the first and second parts, respectively, based on a transmission pattern. We employ gOMP to detect active pilots and patterns, LMMSE channel estimation, and single-user polar decoding to recover the transmitted messages. Numerical examples demonstrate that our proposed scheme offers a higher energy efficiency compared to the state-of-the-art for different numbers of receive antennas, and that it can be employed in environments with a very high connection density.

%and it can support the communication of 1400 active users with a moderate energy-per-bit.
%The first part is utilized for pilot transmission and the active users distribute their data symbols to the second part in an ODMA manner.

\end{document}